\documentclass[a4paper]{jpconf}
\usepackage{graphicx}
\newcommand{\be}{\begin{eqnarray}}
\newcommand{\ee}{\end{eqnarray}}
\begin{document}
\title{Conical flow in a medium with variable speed of sound}

\author{ J.Casalderrey-Solana
and ~E.V. Shuryak}

\address{ Department of Physics and Astronomy\\ State University of New York,
     Stony Brook, NY 11794-3800}

\ead{casalder@tonic.physics.sunysb.edu; shuryak@tonic.physics.sunysb.edu}

\begin{abstract}
 In high energy nuclear collisions, QCD jets  deposit a large fraction of
their energy  into the produced matter.  It has been proposed
that as such matter behaves as a liquid 
 with a very small viscosity, a fraction of this energy goes into
 a   collective excitation 
called the ``conical flow'', similar e.g. to
the sonic booms generated by the supersonic planes.
In this work we study the effect of time-dependent speed of sound
on the development of the conical wave. We show that the expansion of matter and the
decrease of $c_s$ leads to an increase of observable manifestations of
the conical flow. We also show that 
if the QCD phase
transition
is of the first order
(and thus with vanishing speed of sound in the mixed phase)
the wave must split into two, with opposite directions.
We then argue that it  is 
not the case experimentally, which supports the conclusion that
the QCD phase transition is $not$ of the first order.
\end{abstract}

\section{Introduction}
A strong suppression of  particle production at high transverse
 momenta
 has been observed at RHIC \cite{{jetquenching}}, and it is
 commonly interpreted as due to parton energy loss ({\em jet quenching}).
 See early papers \cite{early},
more recent developments \cite{Gyulassy_losses,Dok_etal,SZ_dedx} and a short summary
\cite{xnwang_workshop}.

An interesting question is where the deposited energy/momentum may eventually be observed, and in what form.
Since the state produced at RHIC appears to be a liquid-like QGP \cite{sQGP}
with a very short dissipative length 
(small viscosity)
\cite{Derek_visc}, details such as the angular distribution of
emitted gluons should be rapidly forgotten and energy/momentum
be deposited locally. Further evolution would then be described
by relativistic hydrodynamics.
In \cite{CST} (in a brief form the idea was also mentioned in
\cite{Stocker})
 it was suggested 
 that such local deposition of energy and momentum could start a
 ``conical flow'' of matter. The distortion of the
shock waves due to the expansion has been studied in \cite{S_w} and 
an attempt to incorporate more realistic mediums has been performed in 
\cite{Renk} . Discussion of alternative explanations to the large
$p_t$ correlations at 
RHIC can be found in \cite{RM}, \cite{Dremin}, \cite{MWK} and \cite{Vitev}.

The solution of the linearized hydrodynamics equations \cite{CST} provides
a detailed picture of resulting flow of matter. 
 The main direction of flow
 is at the so called Mach angle relative to the jet velocity
\be 
\label{eqn_Mach}
cos\theta_M={\bar  c_s\over v_{jet}}
\ee
where $v_{jet}$ is the jet velocity and $\bar c_s$ is the (time averaged)
speed of sound in the produced matter. 

For RHIC collisions the 
time-weighted  $\bar c_s$ till the freezeout time $\tau$ is estimated to be
 \be \bar c_s^{RHIC}={1/\tau}\int_o^\tau dt c_s(t)\approx .33 \ee
Since for light quark and gluon jets
$v_{jet}\approx c$,
 the conical flow was calculated in \cite{CST} 
to be at about $\theta=1.2$ radian
or at angles of about 70 degrees relative to the jet.
Experimental observations made at RHIC, first by the STAR collaboration
\cite{STAR_peaks} and then by PHENIX \cite{PHENIX_peaks}, indicate a
depletion of correlated particles in the direction of the quenched jet
and a peak with an angular position and shape in agreement
with hydro predictions. 
The issue whether this effect is real or follow from
incorrect flow subtraction
was discussed in detail by B.Cole \cite{PHENIX_peaks}, who
demonstrated (slide 25) 
that the shape of the correlation function is quite
independent of the direction of the jet with respect to the 
flow. Additionally one may observe it to be a real effect
from  the subset of the data, taken at the particular
 angle $\phi=\pi/4$ at which the elliptic flow vanishes.
 These data  show a clear minimum at
$\Delta \phi=\pi$ without subtraction.
 The  maximum of the correlation function seems to be 
at an angle 110-120 degrees,  an angular position  in rough agreement
with expected Mach cone position. The issue will be further clarified by
the data on three-particle correlations 
reported at QM05: unfortunately at the moment their analysis is too
uncertain to comment on them.

  In this paper we focus on phenomena which are induced by variable
(time-dependent) 
  speed
of sound. This quantity, defined as \be c^2=dp/d\epsilon \ee  
via the thermodynamical variables pressure and energy density, is expected to change significantly
during the process of heavy ion collisions. At early stages at
  RHIC
the matter is believed to be  in the form of
quark-gluon plasma  (QGP), and thus
with $c^2_{QGP}\approx 1/3$. The next stage is the so called ``mixed
  phase''
in which the energy density is increasing much more rapidly than
 the  pressure,
so that $c^2$ decreases to the so called ``softest point'' \cite{Hung:1997du},
and then rises again to $c^2_{RG}=.2$ in the hadronic ``resonance
  gas''.

In this paper we will discuss basically two issues. The first is
 our finding that
a gradual change of the sound speed 
affect the velocity amplitude of the wave.
 We apply rather well developed analytical methods based on {\em 
adiabatic invariants} (see textbooks e.g.\cite{landau}) to infer how these changes \footnote{ 
Note also, that this analytic theory  also helps  to
clarify a (somewhat 
 methodical) question of monitoring 
accuracy of a numerical solution. In a stationary time-independent
matter considered so far, this is done via the energy conservation,
not available any more for any equations with time-dependent
parameters such as  $c^2$.} influence the final velocity
of the sound wave, related the to final particles spectra
in realistic conditions of heavy ion collisions.

 It is well known that  for the description of 
the motion of
any wave in a weakly inhomogeneous medium one can use geometric optics 
and
eikonal equations to derive the $phase$ of the wave.
 The $amplitude$ magnitude is a more
tricky question, which we would like
to address with the help of {\em adiabatic
  invariants} \be I=\oint pdq \ee
It is convenient to remind the reader
what its conservation means for  the basic example,
the harmonic oscillator with a slowly variable frequency
$d log\omega(t)/d log t\ll 1$ and/or mass $d log M(t)/d log t\ll 1$.
 The typical momentum and amplitude
scale as 
\be p\sim
\sqrt{M(t) E(t)},\,\, \, q\sim \sqrt{E(t)/\omega^2(t)M(t)}\ee
where $E(t),\omega(t),M(t)$ are slowly variable energy, frequency and mass.
Their product is the adiabatic invariant $I\sim E(t) / \omega(t)$ 
which should remain a constant, independent on time.

Hydrodynamical equations
for sound waves, in momentum representation in coordinates,
also form an oscillator,
and therefore we will use the adiabatic invariant to study the changes on the 
velocity field due to the expansion and variable speed of sound.
We will conclude that both of these phenomena  enhance the sonic boom effect
in QGP.

The second issue is
what will happen if the QCD phase transition is 1-st order,
 so that
there is a truly mixed phase and the  minimal value of $c^2$
is zero. As we will show below, this indeed would produce quite
dramatic changes: all the waves, including conical ones, stop and then
 resume motion with splitting
into two waves moving in opposite directions.
 Some phenomenological arguments at the end of the paper 
would lead us to believe that this in {\em not} what happens
experimentally,
in correlations with the high-$p_t$ jet. If this is confirmed by 
more accurate calculations and if the correlations observed at RHIC are due to
hydrodynamical ``conical flow'', it would mean that 
QCD phase transition {\em can not} be first order.

\section{Changing the conical wave amplitude by slow perturbations}

\subsection{Sound in expanding matter}
  Unfortunately, a
 simple substitution of a variable $c_s(t)$ 
 into the equations of motion for perturbations of a static background
is inconsistent.
 One should instead find a correct non static solution of the
hydrodynamical equations and only then, using this solution as zeroth order,
study first order perturbations such as sound propagation.
The numerical solutions for hydrodynamical equations
have been done by a number of authors: but in all of them
the flow and matter properties depend on several variables
and is too difficult to implement.

 Therefore, in order to study the effects of the variable speed of 
sound we have looked for the simplest example possible, in
which there is a nontrivial
time-dependent expansion
  but still  no spatial coordinates are involved, keeping the problem
homogeneous in space. The only way these goals can be 
achieved is by a Big-Bang 
gravitational
process, in which the space is created dynamically
by gravity. With such space available, the matter
can cool and expand at all spatial points in the same way.
For definiteness, consider a liquid in flat Robertson-Walker metric \cite{WB}:
\be
d\tau^2=dt^2-R(t)^2\left[dr^2+r^2(d\theta^2+sin^2 \theta d \phi)\right]
\ee
where the  parameter $R(t)$ (the instantaneous
Hubble radius of our ``universe'') is treated
as external. This means that one should not consider Einstein equations
of motion for it explicitly, just look at the effect
of this gravity field on matter motion.

The hydrodynamic equations (for
simplicity, for a baryon free) fluid are the same as usual,
but with the derivatives
replaced by covariant derivatives (denoted
by ;)
\be
T^{\mu \nu}_{;\nu}=0
\ee
 The components of the stress tensor for ideal fluid are, as in the
case of Minkowski space:
\be
T^{\mu \nu}=
(e +p)u^{\mu} u^{\nu}-g^{\mu \nu}
\ee 
This equation can be easily
 solved \cite{WB} for a non-floating medium, 
$u^{\mu}=(1,0,0,0)$. It is straightforward to see
 that out of 4 equations the only nontrivial
one is the longitudinal projection 
$u_{\mu}T^{\mu \nu}_{;\nu}$ leading to the equation
of entropy conservation\footnote{Note that this is also a variant
of adiabatic invariant by itself, true if
the matter expansion is
slow in comparison to the microscopic time scale.}. Thus,
\be
\label{S}
\frac{d}{dt}\left(s(t) R^3(t)\right)=0 \Longrightarrow s(t)R(t)^3=S
\ee 
As we already said, R(t) will be a conveniently chosen external parameter,
 that will drive the expansion dynamics.

The next step is
to derive the linearized equations for hydrodynamical waves
 in this background
metric  $\delta T^{\mu \nu}_{;\nu}=0$ that we can express as,
\be
\partial_t\left(\sqrt{(-g)}\delta T^{00} \right)+
 \partial_i \left(\sqrt{(-g)}\delta T^{0i} \right) 
+3\frac{\dot{R}}{R} c^2_s \delta T^{00} \sqrt{(-g)} =0
 \\ \nonumber
\partial_t\left(\sqrt{(-g)}\delta T^{0i} \right)+
\frac{c^2_s}{R^2} \partial_i \left(\sqrt{(-g)}\delta T^{0} \right)
+2\frac{\dot{R}}{R} \delta T^{0i} \sqrt{(-g)} =0 
\ee
Where we have use $c^2_s=dp/de$ in order to relate  $\delta T^{ij}$ to 
$\delta T^{00}$, and $\sqrt{(-g)}=R^3$.
We define
\be
\epsilon=R^4\delta T^{00} 
\\ \nonumber
G^i=R^5 \delta T^{0i}
\ee
 and perform the standard change of variables 
\be
\eta=\int \frac{1}{R} dt
\ee
such that the metric takes a conformally flat form. After some straight forward
algebra one gets the equations of motion for small perturbations:
\be
\label{perturbation}
\partial_{\eta} \epsilon +\partial_i G^i +(3c_s^2-1) \frac {R'}{R} \epsilon=0
\\ \nonumber
\partial_{\eta}G^i + c^2_s \partial_i \epsilon =0
\ee
where here $R'=dR/d\eta$.

In the derivation of (\ref{perturbation}) we have not assumed any
particular expansion law or 
equation of state, just thermodynamic properties.
Given the time dependence of the Hubble radius
$s(t)$
and provided the equation of state for the fluid, we can obtain all the
thermodynamic properties, in particular $c_s$.

In order to get a better feeling about the different contributions in 
(\ref{perturbation}) let us write the second order equation for the 
quantity $\epsilon$
\be
\label{second}
\partial^2_{\eta} \epsilon - c_s^2 \nabla \epsilon 
+ \epsilon\partial_{\eta} \left( (3c_s^2-1)\frac{R'}{R} \right ) + 
(3c_s^2-1)\frac{R'}{R} \partial_{\eta} \epsilon=0.
\ee
 The resulting wave equation has two new terms,
the first changes the speed of wave propagation and the second is 
a rescaling factor.
Note that the corrections to the wave equations vanish for $c_s^2=1/3$ 
(the QGP value) and the rescaling factor 
 actually produces an amplitude growth, if the speed of sound reduces
below the ideal value 
$c^2_s <1/3$. 

Let us Fourier transform the space components and introduce the
following quantities:
\be
\label{M_eq}
\frac{M'}{M}= 
\left( (3c_s^2-1)\frac{R'}{R} \right ) 
\ee
\be
\omega^2=k^2 c^2_s - \partial_{\eta} \left( (3c_s^2-1)\frac{R'}{R} \right ) 
\ee
then, we can write (\ref{second}) as the equation
 for a harmonic oscillator
with (time dependent) mass M and frequency $\omega$
\be
\partial_{\eta} (M\partial_{\eta} \epsilon) +M \omega^2 \epsilon=0
\ee
As is well known, the Hamiltonian for such system can be written in
the oscillator form
\be
\label{H}
H=\frac{1}{2} \frac{p^2}{M}+ \frac{1}{2} M \omega^2 q^2
\ee
with $p=M d\epsilon/d\eta$ and $q=\epsilon$

One can do a similar analysis for the momentum ${\bf G}$. As in
\cite{CST} in order to have propagating modes (sound waves), the
space Fourier transform of ${\bf G}$ should be in the form:
\be
{\bf G} (t,{\bf k})=\hat{k}G_L=i{\bf k}G
\ee
With this requirement we can also write a harmonic oscillator kind of equation
for $G$ that looks slightly different from the previous. 
After Fourier 
transform of (\ref{perturbation}) the second order equation for G is
\be
\partial_{\eta} (\bar{M}\partial_{\eta} G) +\bar{M} \bar{\omega}^2 G=0
\ee
with $\bar{M}=M/c_s^2$ and $\bar{\omega}=k c_s$. The Hamiltonian $\bar{H}$
as well as the generalized coordinates $\bar{q}$, $\bar{p}$ are defined in 
an analogous way to the previous case.  
 
\subsection{Adiabatic invariants}
  Adiabatic invariants allow to predict
 properties of open mechanical systems
subject to small (adiabatic changes) in the external parameters.
We will now use those to learn
about the effect of the expansion in the final solutions  as well as to
check our numeric procedures.
After the analogy of the equations for the (space) Fourier modes
with those of the 
 harmonic oscillator with time dependent mass and frequency 
is established, it is rather straightforward.

 The adiabatic condition for any time dependent parameter
$\lambda$ of the equation of motion (M or $\omega$  in our case) is
\be
\label{adb_cond}
\frac{1}{\lambda} \frac{d \lambda}{dt} << \frac{1}{T_{osc}}
\ee
where $T_{osc}$ is the period of oscillation of the solution of the 
equation of motion
of the system considered as if the parameters where constant. In our case
$T_{osc}=2 \pi / \omega$

As stated before, if the parameters change in an adiabatic fashion, the 
adiabatic invariant $I$ remains approximately constant through out the evolution
of the system. For the harmonic oscillator the following identity holds
\be
I=\frac{\langle H \rangle}{\omega}
\ee
where H is the Hamiltonian and the average is taken though one period of oscillation
of $q$ (taken as if both M and $\omega$ where constant and equal to the value
at some time). Thus, the amplitude $A_k$ for an oscillation
$\epsilon(k,t)$ 
is
\be
A_k=\sqrt{\frac{2I}{\omega M}}
\ee
In the same way, one
 can perform the same analysis for the time dependent 
harmonic oscillator for G and obtain for the amplitude of the oscillations
$B_k$:
\be
B_k=\sqrt{\frac{2\bar{I}}{\bar{\omega} \bar{M}}}=
\sqrt{\frac{2\bar{I} c_s}{{k} M}}
\ee

Note, however, that these quantities are related to the components of the 
perturbed stress energy tensor though time dependent quantities. 
In particular we would like to look at the behavior of the perturbed velocity
field $v^i=Ru^i$ (where the factor R recovers the correct dimensions),
related to $\delta T^{0i}=wu^i$, with $w$ the enthalpy $w=e+p=Ts$. Thus, 
we find
\be
v^i=\frac{G^i}{R^4w}=\frac{G^i}{STR}
\ee
We now can relate $TR$ to $M$ remembering that for a baryon free fluid 
$c^2_s=s/T dT/ds$, and by use of equation (\ref{S}) 
\be
\label{Tot}
\frac{T'}{T}=-3c^2_s\frac{R'}{R}=-\left(\frac{M'}{M}-\frac{R'}{R}\right)
\ee
Thus, we can fixed the normalization for M such that
\be
T=\frac{1}{MR}
\ee
 Combining this with the previous expression,
\be
v^i=\frac{G^i M}{S}
\ee
And thus the amplitude of the velocity ($v^i_k$  for a fixed Fourier mode ) 
behaves as
\be
v^i_k=\sqrt{M \bar{\omega}2 \bar{I}}
\ee
Therefore, for a medium with $c^2_s<1/3$, where 
M decreases with increasing R (\ref{M_eq}), 
we conclude that the amplitude of the perturbations of the
velocity field decreases with the expansion. The dependence 
on the speed of sound is two fold, through $\bar{\omega}$ and M.
From (\ref{M_eq}) we see that for a fixed expansion rate and initial value
for the temperature,
M decreases faster for smaller $c_s$. As also $\bar{\omega}$ decreases
in an obvious fashion with $c_s$, the amplitude of the velocity perturbations
also decreases with decreasing speed of sound.

  However, as we will show when we discuss the spectrum, the relevant
quantity for the final production of particles is not the velocity
but the ratio $v^i/T=R v^i/M$. 
Thus, the ratio of the amplitude of the wave (with mode k) to the temperature
behaves as
\be
\frac{v^i_k}{T}=R\sqrt{\frac{\bar{\omega} 2\bar{I}}{M}},
\ee
as in the previous case, for $c^2_s<1/3$, it  $increases$ with the 
expansion. The dependence on 
the speed of sound is more complicated due to the ratio $\bar{\omega}/M$,
where both quantities decrease with $c_s$. So in order to determine
the dependence on $c_s$ we need extra information on the time dependence of
both quantities.

\subsection{Monitoring the numerical solution}
In the next section we will study a case where the adiabatic approximation is not
valid. Then, we will need to solve numerically the system of equations (\ref{perturbation}).
We use the point splitting technique, separating (\ref{perturbation}) in a wave equation
and a rescaling part. For the wave equation we use the  MacCormack's technique \cite{And},
and for the rescaling factor a first order Runge-Kutta update.

The adiabatic invariant provides us with a way to check 
our numerical procedure. In the static medium case,
 one can use the energy conservation to monitor the
solution. Now that check is not available and, thus, we need to look
for some other (approximate) constant of motion, that is the adiabatic 
invariant.  

We will illustrate our check in the simplest possible case (after $c_s^2=1/3$), 
that of constant $c_s$ and constant frequency of oscillation mode by setting
\be
R=e^{\alpha \eta}
\ee
with arbitrary  $\alpha$ . With this prescription, only M depends
on $\eta$ in the following way
\be
M=R^{\left(3c^2_s-1\right)}
\ee
The adiabatic condition for this system reads 
\be
\frac{M'}{M}<<\frac{ \omega}{2\pi} \rightarrow 
\gamma=2 \pi \frac{3 c^2_s-1}{c_s k} \alpha <<1 
\ee

\begin{figure}[t] 
\includegraphics[width=7cm]{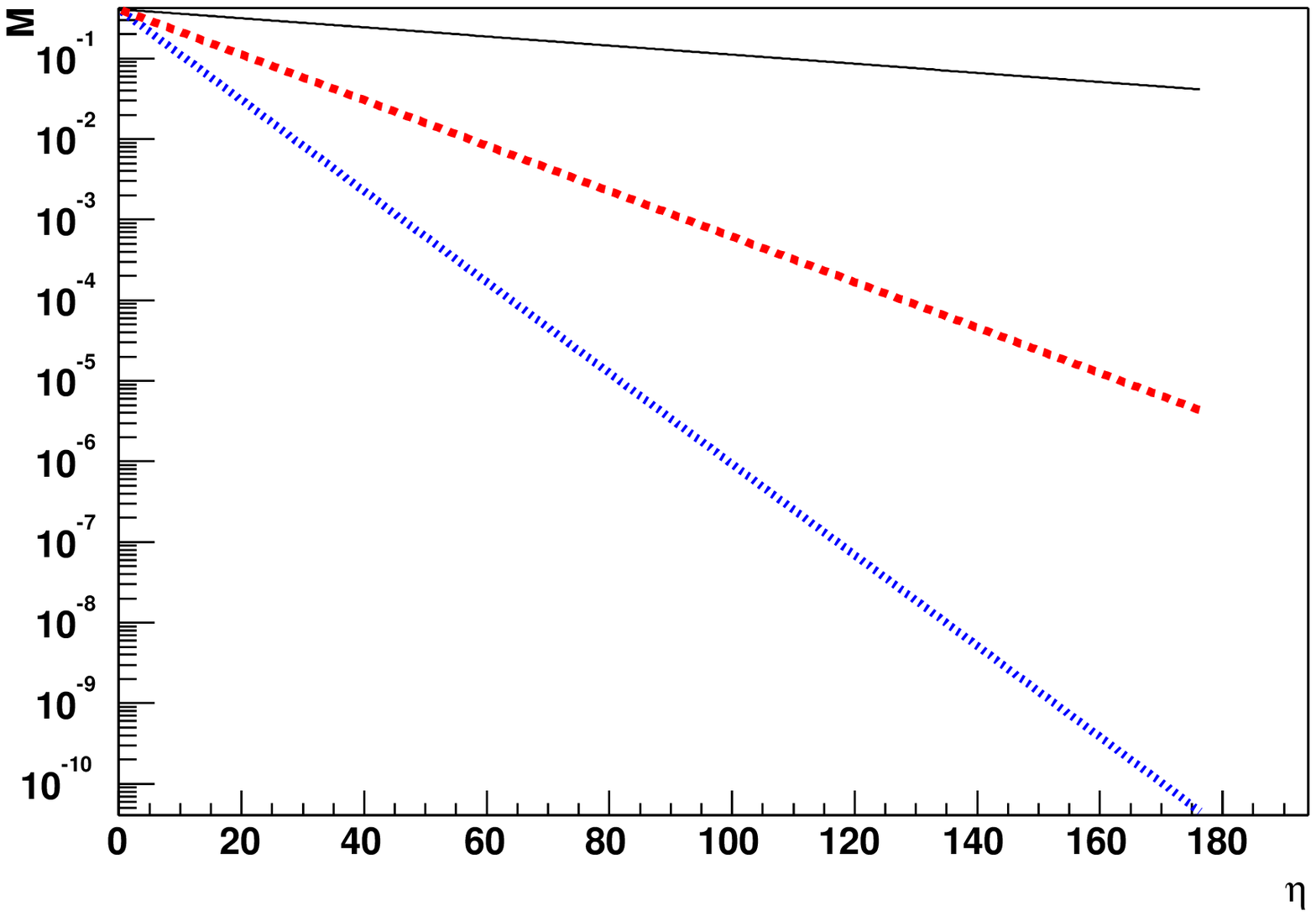}
\includegraphics[width=7cm]{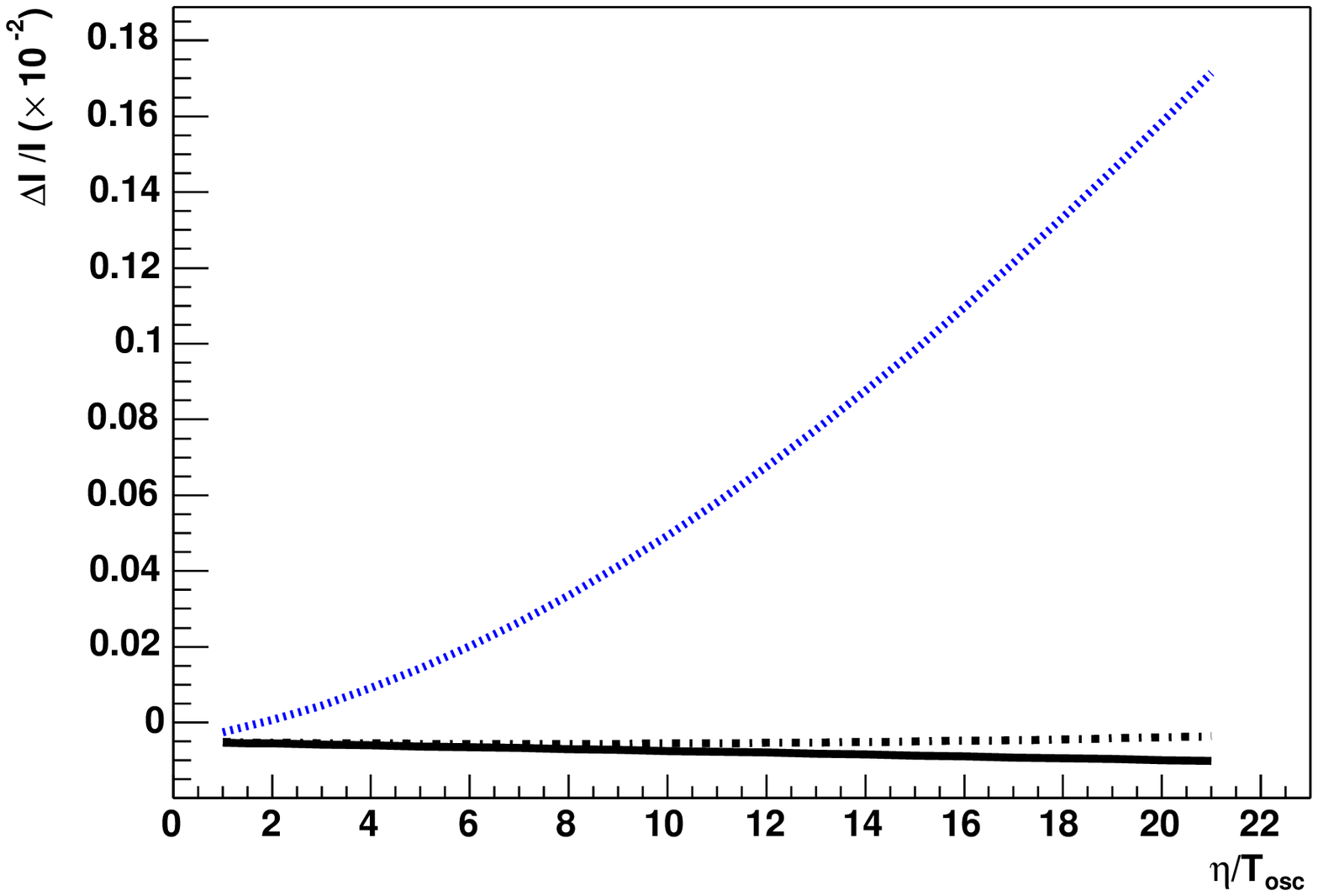}
\caption{
\label{numchec}
Check of our numerical procedure for the simplest possible case (see text)
(a) time dependent mass M for the k=2 mode
(b) Relative change of the adiabatic invariant $\Delta I /I= (I(\eta)-I(0))/I(0)$ in 20 periods of 
oscillation. In both cases $\gamma=0.1,~0.5,~1$ for solid, dashed and dotted lines
respectively. 
}

\end{figure}

In Fig \ref{numchec} we plot the relative change of the adiabatic invariant 
($\Delta I /I= (I(\eta)-I(0))/I(0) $)
for a particular Fourier mode (k=2)
for different values for the coefficient $\gamma$. In practice, we fixed 
the value of $3c^2_s=0.5$ and changed $\alpha$. In Fig \ref{numchec} a
we show, the change of M for different choice of parameters. 
In Fig \ref{numchec} b we show the change of I. 
Even though the mass changes by orders of magnitude, our numerically 
calculated adiabatic invariant remains constant on the level
of 0.01 \%. Even for the values
of $\gamma=1$ the total change of the adiabatic invariant in the
time considered (20 $T_{osc}$) is very small.

\subsection {The wave amplitude and the observed spectrum}
 The spectrum of particles
produced by  hydrodynamical velocity and pressure fields 
is determined at the freeze out. The standard way to calculate it is
the Cooper-Fry prescription, and we discuss how it works in the case
we discussed, with cosmological-style expansion. Although it is not
identical
to experimental conditions, with flow fields depending on spatial
coordinates, one can still get 
some lesson from this example.

The simplest way to obtain a particle spectrum from the hydro calculation
is to consider that the system behaves hydrodynamically till some freeze out
happens and after that particles (created through a thermal distribution) free
stream. If the case of flat space, when the spectrum is calculated though
the Cooper-Fry prescription at some time, the free streaming stage does not
change the invariant spectrum. 

However this is not the case in our curved expanding space because
particle momenta continue to change  in the expanding Robertson Walker space
(as an illustration,
the cosmic microwave background has a Boltzmann distribution, but the temperature
of the distribution is time dependent). Thus, if we would keep the radius
of our ``universe'' to expand for ever, the spectrum of particles 
we calculate after freeze out will continuously change, dropping the measured
temperature.

In order to obtain an answer that resembles more the experimental 
conditions one may  consider the 
``universe'' which stops expanding after the freezeout. Thus, the spectrum remains constant when
the particles fly out of the collision region. As in \cite{CST} 
we will consider fixed proper time freezeout, for which

\be
\frac{dN}{R_{f}^3d^3p}=\int R_f^3d^3x \frac{1}{e^{p^{\mu}u_{\mu}/T_f} \pm 1}
\ee

Where $R_{f}$ is the value of the universe size at freeze out (and constant 
afterwords as an assumption). This factor has to be included to get the
correct dimensions. The exponent of the thermal distribution also
includes such factor. For a small perturbation, it will be
\be
p^{\mu}u_{\mu}=E-R_f p^i R_f u^i 
\ee
From here we observe that for a fixed momentum ($R_t p^i$) the relevant
quantity that sets the correlations in the spectrum is $v/T$. We
have shown already that this combination of fields increases with time.
Thus, the effect of a small disturbance at early times, when the medium
is more dense, increase as the system expands and becomes more dilute,
as the amplitude of the relevant hydrodynamic field grows.

\section{Vanishing speed of sound and reflected waves}
As stated in the introduction, the speed of sound 
reaches a minimum in the mixed phase. 
If the system experiences the first order phase transition, the speed
of sound in the 
mixed phase
should vanish, because
the pressure does not change during this stage while the energy density does.
In this section we study the 
consequences of a vanishing speed of sound for the sound propagation.

Let us start by noting that when the speed of sound vanishes, the adiabatic
approximation explored in the previous section can no longer be applied. In fact,
in such a case, the logarithmic derivatives of $c_s$ diverges and thus,
 (\ref{adb_cond}) does not hold. Thus, we will need
to solve numerically the equations (\ref{perturbation}) with the procedure checked
in the previous section in order to investigate  this effect.  

In order to get somewhat realistic estimates of the effects, we will model
both the expansion rate and the speed of sound according to the physical 
situation at RHIC. 
Inspired in the boost invariant scenario, 
we will chose R(t) such that the entropy density behaves as:
\be
s(t)=s_0\frac{t_0}{t_0+t} \Longrightarrow R(t)=R_0 \left( \frac{t_0+t}{t_0} \right)^{1/3}
\ee	
Where $t_0=0.5 fm$ and $R_0=6 fm$. The value of $s_0$ will be specified later.

\begin{figure}[t] 
\includegraphics[width=7cm]{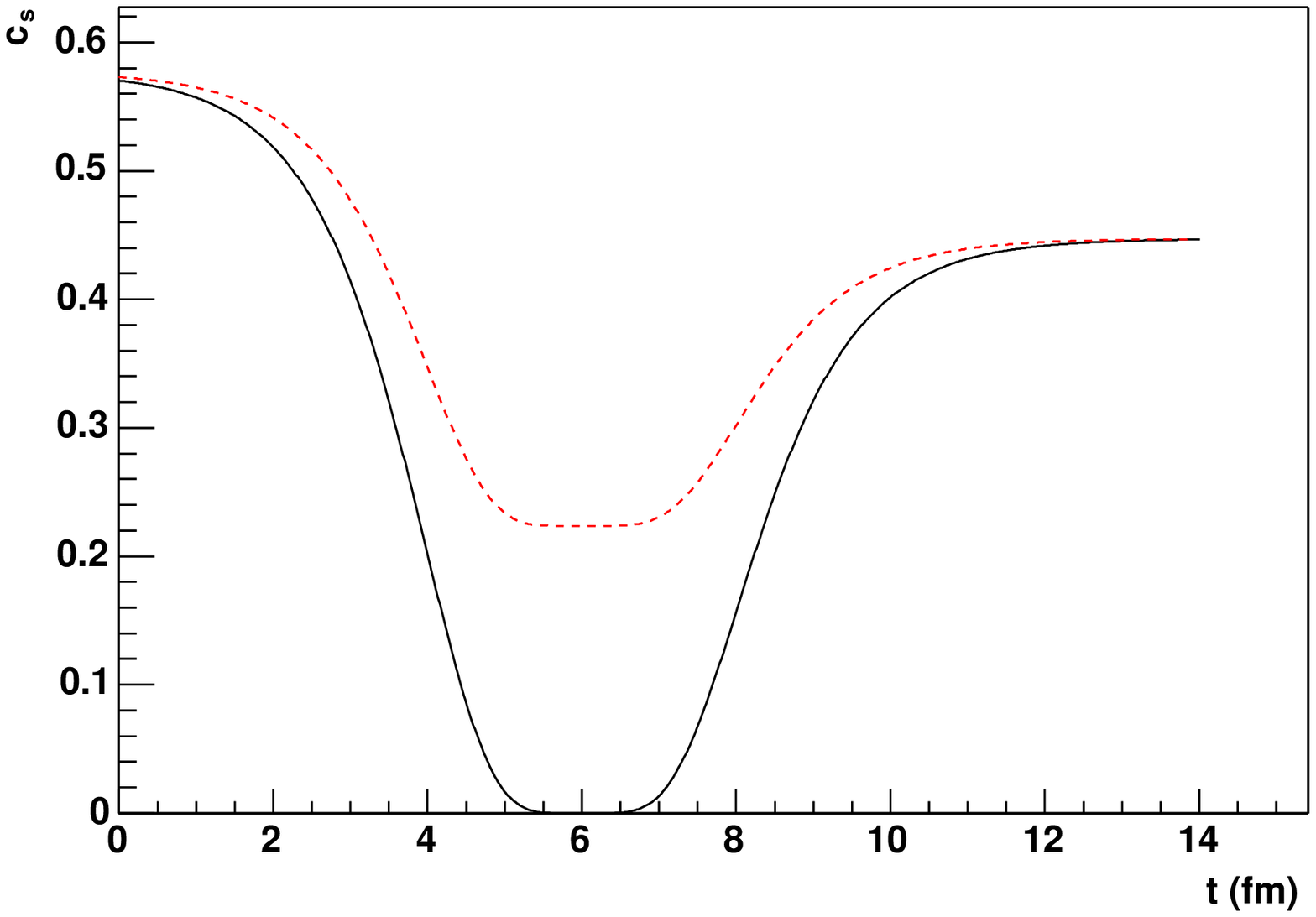}
\includegraphics[width=7cm]{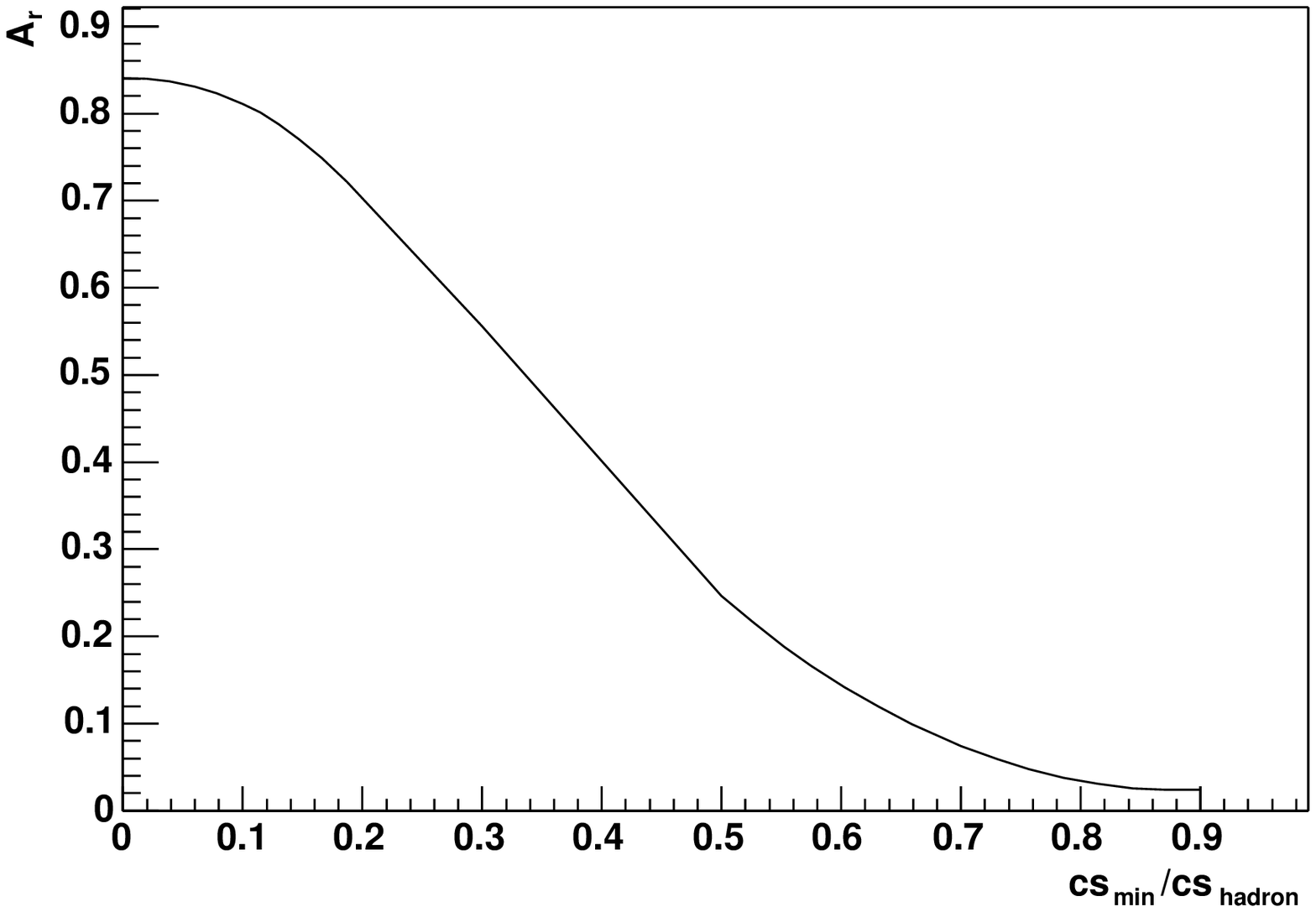}
\caption{
\label{csch}
(a) $c_s$ as a function of t. 
The two curves are different interpolations with different 
minimal at the mixed phase (0 (solid) and $1/2 \sqrt{2}$ (dashed))
(b) Ratio of  amplitudes of the reflected and transmitted waves 
as a function of the minimal 
$c_s$ in units of $c_s$ of the hadron gas 
}
\end{figure}

We now discuss the function $c_s(t)$.
Numerical solutions of hydrodynamics as in \cite{TLS} use values
of the speed of sound that are discontinuous at the beginning of the three phases. 
Here, however, we  study parametrization of the 
speed of sound that interpolate smoothly
between the three stages of matter at RHIC. The times for the interpolation
are also inspired in hydrodynamic solutions for RHIC \cite{TLS}. 
Following these calculations we will assume that each of the three stages 
have the same
time duration of 4 fm. Finally, in order to investigate the effect of the
first order phase transition on the sound waves, 
we will leave the minimum value of the speed of sound in the
mixed phase as a free parameter and we study the propagation of the waves
for different values of this minimum. 
(see Fig. \ref{csch} a) ). Once the function $c_s(t)$ is determined, we can solve
(\ref{Tot}) to obtain the temperature, imposing that at the time where $c_s(t)$ has
a minimum ($t_{min}=6$ fm in our case) $T(t_{min})=T_c$. This determines uniquely the temperature, and in particular
the value $T(0)$. We use this value of the temperature and the equation of state for
an ideal QGP in order to determine $s_0$.

Once all the parameters are specified, we proceed to numerically solve (\ref{perturbation}).
We will study the propagation of a spherical wave started from a perturbation at time t=0.
Let us start our discussion with the case of vanishing minimum $c_s$ (first order phase
transition). In this case, the spherical wave 
will split into two at the mixed phase, one propagating outwards (transmitted wave) and the second
 inwards (reflected) towards the origin. A simple physical explanation
 of this effect is 
as follows: when the speed of sound
vanishes, all the information about the directionality of the wave disappears and, thus, when the 
speed of sound changes again to a finite value the disturbance splits into two, as it can 
propagate both inwards and outwards. Finally, the spherical wave propagating inwards eventually
bounces back when it arrives at the origin, leading to a second
outward moving spherical wave. This is observed in numerical solution
as a double peaked structure of the 
spherical wave profile at sufficient late times.

The appearance of the reflected wave happens not only when the speed
of sound vanishes. We studied the spherical wave profile for different 
values of the minimum. This is best done at sufficiently late time, when
the two waves (transmitted and reflected) propagate in the same direction. 
To illustrate the importance of the reflected wave, we study the amplitudes
of the maximums in the wave profile for both waves at a late time.
 (In order to get rid of a trivial
$1/r$ decay of the wave, we multiply them by  the position of the maximum r.)
 Thus we define the relative amplitude: 
\be
A_r=\frac{r\epsilon_{reflected}}{r\epsilon_{transmitted}}
\ee
where the values are  to be taken at the maxima of the disturbances.
We plot this quantity in Fig \ref{csch} b). We observe that, as claimed, the
reflected wave appears for non vanishing values of the minimal speed 
of sound. However the relative importance of the reflected wave decreases as
the minimal value increases (as expected)

Even though the particular dependence of the relative amplitude with the
minimal speed of sound, depends on our 
assumed parametrization $c_s(t)$, the general phenomenon
is independent of the parametrization. As long as the speed of sounds
vanishes at the mixed phase, we will obtain a reflected wave for the
spherical disturbance.

Finally, for the cases where the minimal speed of sound does not vanish (as in the 
dashed curved in Fig.\ref{csch} a ), we can use the adiabatic invariant to
learn about the amplitude dependence of the disturbance. As we have specified
a particular time dependence of $R(t)$ and $c_s(t)$, we can complete the analysis
of the previous section for this particular case. By solving (\ref{M_eq}) we
find that for our parametrization of the speed of sound with minimum ranging 
from $0.5 \sqrt{0.2}$ to $0.9 \sqrt{0.2}$, the effect of the overall dropping speed 
of sound increases
the ratio $v/T$ by 10\%. As before, the expansion of the medium also
induces an amplitude increase. For our particular expansion rate, setting the freeze out
time to $t_{f}\approx 14 fm$, we obtain for the same quantity a factor of 3 increase (as 
the system becomes three times larger). These changes are very
important experimentally, as the ratio $v/T$ enters in the exponent of
the Cooper-Fry formula for the observed spectra.

\subsection{Comparison to experimental correlation functions }
Now we discuss where  such reflected waves, if produced, will be located
in the conical flow induced by high energy jets. The cone arises
from the superposition of spherical waves produced along the path of
the jet in the medium. All those disturbances produced prior to the 
mixed phase will encounter a time of minimum speed of sound. Following
our previous discussion, if this minimum is small enough, all those 
waves will split into transmitted and reflected waves. After the
superposition, given sufficient time,
the reflected waves will generate a second cone. So, provided the observation time
is late enough as for all the reflected wave to bounce back, one would find 
two cones with different opening angle propagating in the same
direction. 

However if the observation time is too close to emission time, one 
would find the the second cone moving inwards (or opposite direction to the jet).
At RHIC the timing of the expansion is such that the three main stages --
the QGP, the mixed phase and the hadronic phase -- take roughly
comparable period of proper time, about 4-5 fm/c each.
The speed of sound at the QGP phase $c_s^{QGP}=1/\sqrt{3}$ 
is larger than
that in the final hadronic phase   $c_s^{H}=\sqrt{.2}$. Therefore,
the distance the conical wave would propagate in QGP is larger than
that in hadronic phase, and thus the reflected wave $cannot$ possibly reach
the origin (for spherical wave, the original axes for conical one).

\begin{figure}[t] 
\begin{center}
\includegraphics[width=7cm]{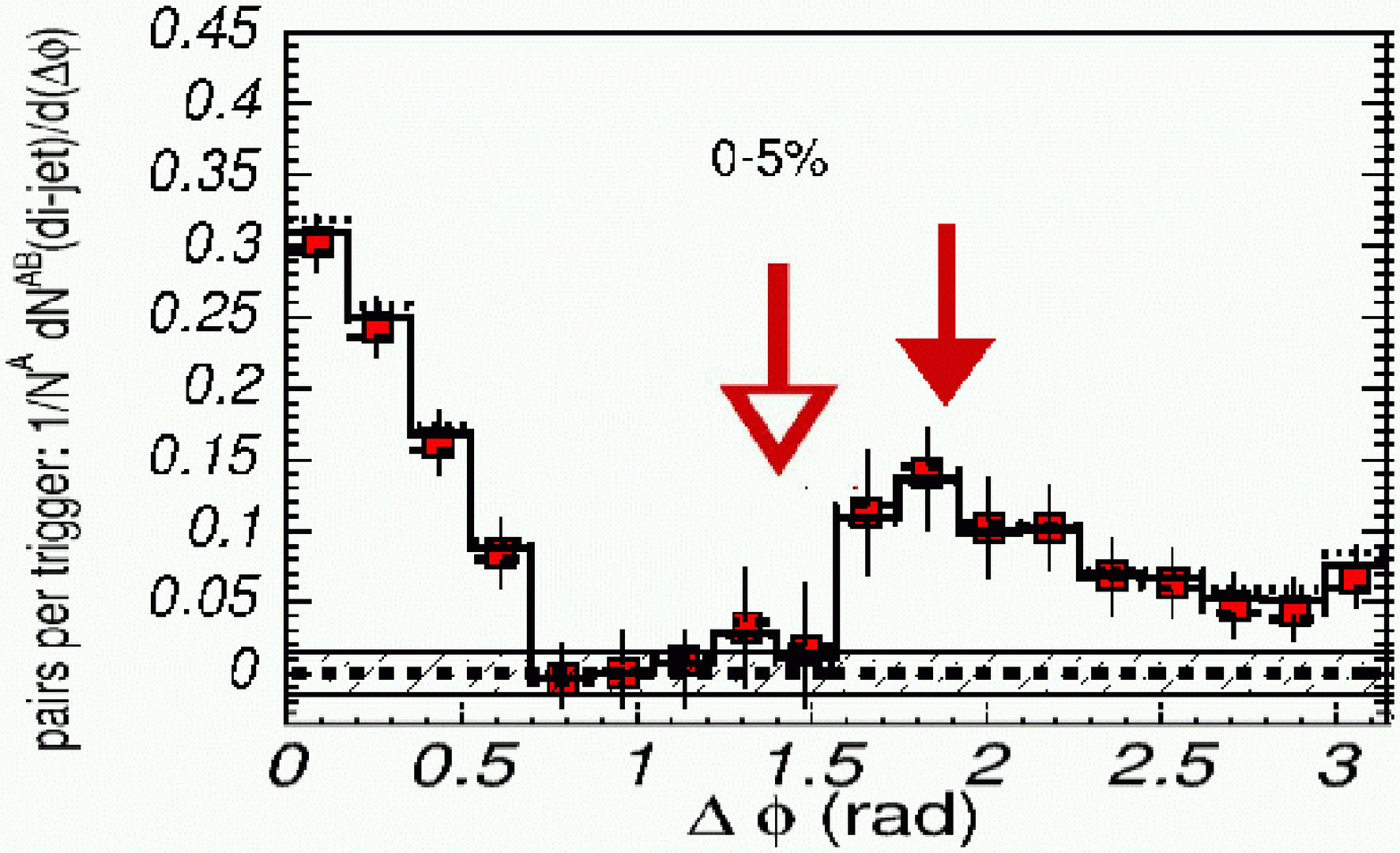}
\includegraphics[width=7cm]{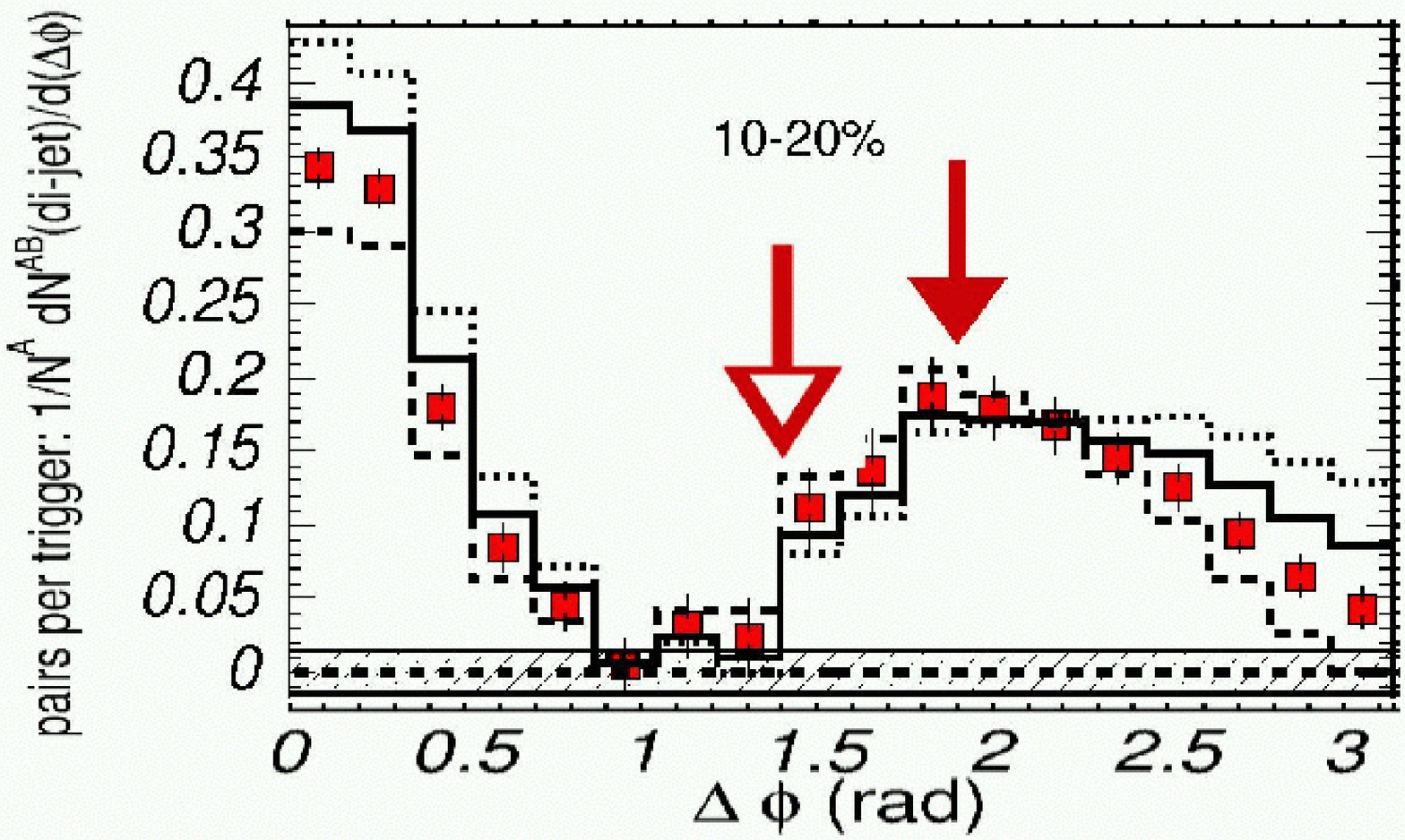}
\end{center}
\caption{
\label{exp_corr}
Azimuthal dihadron distributions normalized per trigger particle measured
by PHENIX \cite{PHENIX_peaks} for two different centralities 
($2.5 < p_T^{trigger} < 4 GeV$,$1 < p_T^{associated} < 2.5 GeV)$. 
The filled arrow indicates the position of the Mach cone. The empty
arrow our estimate for the position where
the cone from the reflected wave should appear. 
}
\end{figure} 

What this conclusion implies experimentally, is that $if$ the QCD phase transition is
 first order and the original wave gets split into direct and
reflected waves, the freezeout will find the reflected wave on its way
 $to$ the origin, the opposite to the Mach direction of the
direct wave.  In terms of two particle 
azimuthal distribution it means the appearance of a second peak for
$\Delta \phi < \pi/2$. Even though detailed numerical simulations
are needed to give the correct position of the peak, we give here
a simple estimate for its location, based on geometrical optics.

The disturbance generated at the origin travels a distance AB
\be
AB=\int^{t_M} _0 dt c_s(t)
\ee
before the minimum of $c_s$ is reached and the wave splits into two.
During the mixed phase the disturbance does not propagate, but after
the hadronic stage starts, the waves propagates backwards a distance
\be
BC=\int^{t_f}_{2t_M} dt c_s(t)
\ee
Where $t_f$ is the freeze out time and we have assumed that
the mixed QGP phase and the mixed phase have the same time duration.
Thus, the relevant distance
traveled by the wave is AC=AB-BC. The relevant longitudinal distance
is not the distance traveled by the jet, but the distance traveled 
till the minimum of $c_s$, $t_M$ (as after that time there is not more
``emission''  of reflected waves). Thus, the effective angle of emission is
\be
cos (\theta_e)=AC/t_M \Longrightarrow \theta_e \approx 1.4~ rad
\ee
where we used for AB and BC fixed $c_s$ values of $1/\sqrt{3}$ and $\sqrt{0.2}$ 
respectively.

 In Fig. \ref{exp_corr}  we show samples of experimental correlation function as obtained
by PHENIX \cite{PHENIX_peaks}. 
The peak around 1.9 rad  (indicated by the 
filled arrow) corresponds to the Mach direction; this is the main
effect attributed to a conical flow. The  reflected wave should appear in 
the region $\Delta \phi< \pi/2$, and the empty arrow shows the estimated
place where the corresponding reflected peak should
 appear. Fig \ref{exp_corr} a) shows the most central collisions,
does not show any nonzero signal at that angle. Fig \ref{exp_corr} b), at
higher centrality, has a nonzero correlation function there, but  
it seems likely to be just a slope
of  a much broader peak.

 We thus argue that there are no indications for enhanced
correlations at the expected angle of the reflected wave,
and thus  the deconfining
phase transition cannot be of the first order.

\section{Summary and discussion}
In this paper we have studied the effect of a variable speed of sound
in the propagation of sound waves. We have considered for simplicity
an expanding Universe
with a flat Robertson-Walker metric, that allows 
us to introduce dynamical expansion depending on proper time only.

 We have use the adiabatic approximation to investigate 
the effect of the expansion on the waves, concluding that even though
the velocity field decreases with the expansion, the ratio of 
the velocity to the temperature increases. This ratio enters
the exponent of the Cooper-Fry formula, and thus the observed effects in
particle spectra may be significantly enhanced, the more so the larger
is the particle momenta. We have shown that within the model studied 
based on the Robertson-Walker metric we obtain a 10 \% increase
of the ratio $v/T$ due to the change of $c_s$ and a factor 3 increase
of this ratio
due to the expansion. Even though this values depend on the particular model
studied we expect a similar effect to take place in fireball expansion.

We have also studied the effect of a vanishing speed of sound, where the 
adiabatic approximation cannot be used. This is supposed to happen in the 
mixed phase $if$ the phase transition in QCD is first order. We have shown 
that this must lead to the appearance of reflected waves, the superposition
of which leads to a second cone in the flow field. Given the timing
of the different stages at RHIC the second cone should be moving 
inwards at the time of freezeout.  We concluded that
 the first order phase transition should lead to a peak 
at a relative angle $\Delta \phi \approx 1.4 ~rad$. The experimental 
correlation functions do not show any enhancement at this direction,
which we think  indicates that the phase transition in QCD 
{\em cannot be of the first order}. Needless to say,
 more theoretical and experimental work is needed before finalizing
 this important conclusion.

\vskip 1cm
This work was partially supported by the 
 Department of Energy (U.S.A.) under grants DE-FG02-88ER40388
and DE-FG03-97ER4014.


\newpage

\begin{thebibliography}{99}

\bibitem{jetquenching} Adcox K et al (PHENIX) 2002 {\it Phys. Rev. Lett.} {\bf 88} 022301
\nonum
Adler C et al (STAR) 2002  {\it Phys. Rev. Lett.} {\bf 89} 202301
\nonum
Adler S S et al (PHENIX) 2003 {\it Phys. Rev. Lett.} {\bf 91} 072301
\nonum
Adams J et al (STAR) 2003 {\it Phys. Rev. Lett.} {\bf 91}, 172302

\bibitem{early}
Bjorken J D ,
FERMILAB-PUB-82-059-THY
\nonum
Appel D A 1986 {\it Phys. Rev.} D {\bf 33} 717
\nonum
Blaizot J P and McLerran L D 1986
{\it Phys. Rev.} D {\bf 34} 2739
\bibitem{Gyulassy_losses}
Gyulassy M and Plumer M 1990
{\it Phys. Lett.} B {\bf 243} 432
\nonum
Wang X N, Gyulassy M and Plumer M, 1995
{\it Phys. Rev.} D {\bf 51} 3436
\nonum
Fai G, Barnafoldi G G, M.~Gyulassy, Levai P, Papp G, 
Vitev I and Zhang Y 2001 {\it Preprint} hep-ph/0111211

\bibitem{Dok_etal}
Baier R, Dokshitzer Y L, Peigne S and Schiff D 1995
{\it Phys. Lett.} B  {\bf 345} 277
\nonum
Baier R, Dokshitzer Y L, Mueller A H and Schiff D 2001
{\it JHEP} {\bf 0109} 033

\bibitem{SZ_dedx}
Shuryak E V and Zahed I 2004
{\it Preprint}  hep-ph/0406100

\bibitem{xnwang_workshop} 
Wang X N, 2004 {\it Preprint} nucl-th/0405017


\bibitem{sQGP}
  E.~Shuryak,
  ``Why does the quark gluon plasma at RHIC behave as a nearly ideal fluid?,''
  Prog.\ Part.\ Nucl.\ Phys.\  {\bf 53}, 273 (2004)
  [arXiv:hep-ph/0312227].


\bibitem{Derek_visc} Teaney D 2003 {\it Phys.Rev.} C {\bf 68}
   034913


\bibitem{CST}J.~Casalderrey-Solana, E.~V.~Shuryak and D.~Teaney,
  arXiv:hep-ph/0411315.
To be published in Journal of Physics G: in
Proccedings of Workshop on Correlations and Fluctuations in Relativistic Nuclear Collisions, MIT, April 21-23, 2005

\bibitem{Stocker} Stocker H 2004, nucl-th/0406018

\bibitem{S_w} L. M. Satarov, H. Stoecker, I. N. Mishustin, {\it Phys. Lett.} {\bf B } 627 64,2005

\bibitem{Renk} T. Renk, J. Ruppert,  hep-ph/0509036 

\bibitem{RM} J. Ruppert and B. Muller, {\it Phys. Lett.} {\bf B} 618 123,2005

\bibitem{Dremin} I. M. Dremin,  hep-ph/0507167 

\bibitem{MWK} V. Koch, A. Majumder, X. N. Wang, nucl-th/0507063 

\bibitem{Vitev} I. Vitev, hep-ph/0501255 

\bibitem{STAR_peaks}
 Fuqiang Wang for STAR collaboration, (Quark Matter 2004),
 {\it J.Phys.G} 30 S1299-S1304,2004;
 {\it Preprint} nucl-ex/0404010, also Proceedings of the 
MIT workshop 2005.
\bibitem{PHENIX_peaks} Modifications to di-jet hadron pair
 correlations in Au+Au collisions at S(NN)**(1/2) = 200-GEV. PHENIX
 coll,  nucl-ex/0507032, submitted to PRL; 
 B.Jacak (for PHENIX), Int.conf. on Physics and
 Astropysics of QGP, Calcutta 2005,nucl-ex/0508036 ; 
 B.Cole, Quark Matter 05, August 2005 

\bibitem{Hung:1997du}
  C.~M.~Hung and E.~V.~Shuryak,
  Phys.\ Rev.\ C {\bf 57}, 1891 (1998)
  [arXiv:hep-ph/9709264].

\bibitem{landau} L. D. Landau, E. M. Lifshitz, {\it Mechanics}, Pergamon Press, 1976 

\bibitem{WB} Weinberg S, Gravitation and cosmology: principles and applications of the general theory of relativity. New York, Wiley (1972)  

\bibitem{And} J. D. Anderson, {\it Computational Fluid Dynamics}, McGraw-Hill Inc. (1995)

\bibitem{TLS} D. Teaney, J Lauret, E. V.  Shuryak, arXiv:nucl-th/0110037

\end{thebibliography}
\end{document}